\documentclass[twocolumn,
               showpacs,
               preprintnumbers,
               prd,
               superscriptaddress,
               10pt,
               notitlepage,
               aps]{revtex4-1}

\usepackage{amssymb}
\usepackage{amsmath}
\usepackage{amsfonts}
\usepackage{amsbsy}
\usepackage{dsfont}
\usepackage{mathrsfs}
\usepackage{bm}
\usepackage[inline]{enumitem}
\usepackage{float}

\usepackage[linktocpage,breaklinks]{hyperref}
\usepackage[usenames,dvipsnames]{color}

\usepackage{epsfig}
\usepackage{graphicx}
\usepackage{epstopdf}

\usepackage[linktocpage]{hyperref}
\usepackage[usenames,dvipsnames]{color}
\usepackage{empheq}
\hypersetup{colorlinks=true,
            citecolor=NavyBlue,
            linkcolor=NavyBlue,
            urlcolor=NavyBlue}

\definecolor{darkgreen}{rgb}{0.2,0.7,0.2}

\newcommand{\rd}{{\rm d}}

\def\be{\begin{equation}}
\def\ee{\end{equation}}

\newcommand{\note}[1]{\text{\tiny{#1}}}

\newcommand{\bfi}[1]{\textit{#1$-$}}

\begin{document}

\title{Spin-induced scalarized black holes}

\author{Carlos A. R. Herdeiro}
\affiliation{Departamento de Matem\'atica da Universidade de Aveiro and CIDMA,
Campus de Santiago, 3810-183 Aveiro, Portugal}

\author{Eugen Radu}
\affiliation{Departamento de Matem\'atica da Universidade de Aveiro and CIDMA,
Campus de Santiago, 3810-183 Aveiro, Portugal}

\author{Hector O. Silva}
\affiliation{Max Planck Institute for Gravitational Physics (Albert Einstein Institute),
Am M\"uhlenberg 1, Potsdam 14476, Germany}
\affiliation{Illinois Center for Advanced Studies of the Universe \&
Department of Physics, University of Illinois at Urbana-Champaign,
Urbana, Illinois 61801, USA}

\author{Thomas P. Sotiriou}
\affiliation{School of Mathematical Sciences \& School of Physics and Astronomy,
University of Nottingham, University Park, Nottingham, NG7 2RD, United Kingdom}

\author{Nicol\'as Yunes}
\affiliation{Illinois Center for Advanced Studies of the Universe \&
Department of Physics, University of Illinois at Urbana-Champaign,
Urbana, Illinois 61801, USA}

\begin{abstract}
It was recently shown that a scalar field suitably coupled to the
Gauss-Bonnet invariant $\mathcal{G}$ can undergo a spin-induced linear
tachyonic instability near a Kerr black hole.
This instability appears only once the dimensionless spin $j$ is sufficiently
large, that is, $j \gtrsim 0.5$. A tachyonic instability is the hallmark of
spontaneous scalarization.
Focusing, for illustrative purposes, on a class of theories that do exhibit
this instability, we show that stationary, rotating black hole solutions do
indeed have scalar hair once the spin-induced instability threshold is
exceeded, while black holes that lie below the threshold are described by the
Kerr solution. Our results provide strong support for spin-induced black hole
scalarization.

\end{abstract}

\maketitle

\bfi{Introduction.}
Black holes (BHs) are central players in astrophysics. The recent detections of
gravitational waves~\cite{Abbott:2016blz,LIGOScientific:2018jsj} and the first
BH imaging~\cite{Akiyama:2019cqa} have consolidated the evidence for their
physical reality.
Under the leading paradigm, astrophysical BHs are described by the Kerr
metric~\cite{Kerr:1963ud}.
Astonishingly, this hypothesis entails this macroscopic class of
objects, ranging 10 orders of magnitude in mass, having only 2 (macroscopic)
degrees of freedom: mass $M$ and spin $J$.

A tantalizing possibility beyond the Kerr hypothesis is that astrophysical BHs
are not described by the Kerr metric only in certain regimes.
For instance, if ultralight bosonic fields exist, e.g., as dark matter,
they may undergo a superradiant instability near Kerr
BHs~\cite{Brito:2015oca}, forming a bosonic cloud~\cite{Arvanitaki:2010sy},
which, in  some cases, leads to  new stationary
BHs~\cite{Herdeiro:2014goa,Herdeiro:2016tmi,East:2017ovw}.
The instability, however, is  only efficient for a range of BH masses
determined by the ultralight field's
mass~\cite{Dolan:2012yt,Witek:2012tr,Dolan:2018dqv}.

The prospect of such elusive non-Kerr BHs takes a different guise in gravity
theories that allow BH scalarization~\cite{Silva:2017uqg,Doneva:2017bvd}.
Theories that fashion a coupling between a scalar and the Gauss-Bonnet
invariant can exhibit a tachyonic instability near BHs when the BH spin exceeds
a certain threshold~\cite{Dima:2020yac}.
Interestingly, crossing that threshold also allows these models to circumvent a
known no-hair theorem \cite{Silva:2017uqg,Dima:2020yac}. Hence, one expects
that stationary BHs in these models will exhibit spin-induced scalar hair only
when they are rapidly spinning. As we show below, this is indeed the case.

\bfi{Spontaneous scalarization.}
This effect was first discussed  by Damour and Esposito-Far\`ese
(DEF)~\cite{Damour:1993hw,Damour:1996ke} for compact stars in scalar-tensor
theories of gravity.
The DEF model demonstrated that, if suitably coupled to gravity, a
new field could go undetected in weak field tests of general relativity (GR)
and still have an influence in the strong field of neutron stars, providing strong motivation
for GR tests with binary pulsars. Indeed, the latter have severely
constrained the DEF model~\cite{Freire:2012mg,Shao:2017gwu,Anderson:2019eay},
although the constraints can be evaded if the field is massive~\cite{Ramazanoglu:2016kul}.

In the DEF model (massless or massive), scalarization happens only for stars and
does not affect BHs~\footnote{Matter in the vicinity of BHs
could lead to scalarization in
principle~\cite{Cardoso:2013fwa,Cardoso:2013opa}, but it is not clear if
astrophysical environments provide sufficiently high matter densities for the
effect to take place.}, since, in fact, the model is covered by no-hair theorems
\cite{Hawking:1972qk,Sotiriou:2011dz,Sotiriou:2014pfa}.
However, it was recently shown that scalar-tensor theories that exhibit
BH scalarization do exist~\cite{Silva:2017uqg,Doneva:2017bvd}.
Consider a scalar-Gauss-Bonnet (sGB) theory with action
\begin{equation}
\label{eq:gbaction}
S=
\frac{1}{16\pi}\int \rd ^4x \sqrt{-g} \left[  R - 2\partial_\mu \phi\partial^\mu \phi
 +   \lambda^2 f(\phi) \, \mathcal{G}   \right] \ ,
\end{equation}
where $\mathcal{G}\equiv R^{\mu\nu\rho\sigma}R_{\mu\nu\rho\sigma}-4~R^{\mu\nu}R_{\mu\nu}+R^2\,$
is the Gauss-Bonnet invariant, $\lambda$ (with units of length) determines the
coupling strength between scalar field and $\mathcal{G}$ and $f$ is a
dimensionless function of the scalar field $\phi$. (We work with units where
$G  = 1 = c$).
If $f'(\phi_0)=0$~\footnote{Prime denotes derivative with respect to the argument.},
for some constant $\phi_0$, GR vacuum solutions, together with
$\phi=\phi_0=$~constant, are admissible solutions of the field equations
derived from Eq.~\eqref{eq:gbaction}.
This condition excludes the dilatonic ($f \propto \exp(\phi)$) and
shift-symmetric ($f \propto \phi$) subclasses of sGB in which BHs always have
scalar hair~\cite{Kanti:1995vq,Sotiriou:2013qea,Sotiriou:2014pfa,Delgado:2020rev}.
The constant $\phi_0$ solutions are, in fact, unique thanks to a no-hair
theorem~\cite{Silva:2017uqg}, provided that
\begin{equation}\label{eq:HairCondition}
f''(\phi_0)\,\mathcal{G} < 0\,.
\end{equation}
Interestingly, $-\lambda^2 f''(\phi_0)\,\mathcal{G}/4$ is the effective
mass squared for scalar field perturbations around the GR solution, and, in this
sense, the condition in Eq.~\eqref{eq:HairCondition} ensures the absence  of
tachyonic instabilities.

This suggests that scalarization can occur if Eq.~\eqref{eq:HairCondition} is violated.
Indeed, as a simple example consider the choice $f(\phi)=\phi^2/2$.
For $\phi=0$, the Schwarzschild BH is an admissible solution and
$\mathcal{G} = 48 M^2/r^6$, where $M$ is the Arnowitt-Deser-Misner (ADM) mass.
Evaluated on the horizon, the effective mass squared of scalar perturbations is
then $-3 \lambda^2 /(16M^4)$, indicating the possibility that a tachyonic
instability can take place.
In general, the effective mass can be somewhat negative and still have a stable
configuration~\footnote{A famous example, in a different context, is the
Breitenlohner-Freedman bound in AdS spacetimes~\cite{Breitenlohner:1982jf}.},
but the scalar field perturbation will become unstable if the dimensionless
ratio $M/\lambda$ is made sufficiently small.
In practice, if $M/\lambda \lesssim 0.587$, the scalar field will develop a
tachyonic instability, whose end point might be a scalarized BH~\cite{Silva:2017uqg}.

The fact that the onset of scalarization is captured in linear theory allows
one to identify all possible couplings to curvature that can lead to
scalarization~\cite{Andreou:2019ikc} and simplifies the investigation of the
relevant thresholds~\cite{Ventagli:2020rnx}.
It also makes it straightforward to generalize the mechanism to
nongravitational couplings (see, e.g.,~\cite{Herdeiro:2018wub,Fernandes:2019kmh}).
However, the end point of the instability depends on nonlinear
interactions, as these are the ones that eventually quench the linear
instability. For example, although static, spherically symmetric scalarized
BHs exist for both $f(\phi)\propto \phi^2$~\cite{Silva:2017uqg} and
$f(\phi)\propto e^{\phi^2}$~\cite{Doneva:2017bvd}, they have different radial
stability properties \cite{Blazquez-Salcedo:2018jnn}.
This can be attributed to the additional nonlinear interactions between $\phi$
and $\mathcal{G}$ in the second model~\cite{Silva:2018qhn}.
Alternatively, supplementing the simplest choice, $f(\phi)=\phi^2/2$, which already
determines fully the onset of scalarization, with a nonlinear potential for the
scalar also yields radially stable (and entropically preferred) scalarized BHs~\cite{Macedo:2019sem}.

\bfi{BH rotation.}
The effect of rotation on BH scalarization was considered in
Ref.~\cite{Cunha:2019dwb} for the choice
\be
f(\phi)=\frac{\epsilon}{12} \left( 1-e^{-6\phi^2} \right) \,,
\label{coupling}
\ee
and $\epsilon=+1$~\footnote{The constant term is introduced for convenience and
does not contribute to the field equations, as ${\cal G}$ is a total divergence.
For small values of $\phi$ this theory is encompassed under the effective theory
formulated in~\cite{Macedo:2019sem}.}.
It was shown that rotation tends to suppress scalarization. This can be partially
understood in an intuitive manner as follows.
For a Kerr BH in Boyer-Lindquist coordinates $(t,r,\theta,\varphi)$ one has
\begin{equation}\label{eq:gbterm}
\mathcal{G}_\note{Kerr}=\frac{48 M^2}{(r^2+\chi^2)^6}\left(r^6-15r^4\chi^2+15r^2\chi^4-\chi^6\right)\,,
\end{equation}
where $\chi\equiv a\cos\theta$, $a = J / M$ is the Kerr spin (per unit mass)
parameter, where $J$ is the angular momentum.
When $a=0$, one recovers the Schwarzschild metric, where ${\cal G}$ is
positive definite and monotonic in $r$.
For the Kerr metric, as long as the dimensionless spin $j \equiv a/M
\leqslant 0.5$, $\mathcal{G}$ remains positive definite and the spacetime is
said to be ``gravitoelectric dominated''.
However, this is no longer true when $j > 0.5$ and regions of
``gravitomagnetic dominance'' in which $\mathcal{G}$ is negative
arise for some neighborhoods of $r,\theta$~\cite{Cherubini:2003nj}.
Thus, rotation can make the effective mass of the scalar field less negative or
even positive near the horizon for $\epsilon=+1$ and therefore suppress the
effect of scalarization.

The focus on $\epsilon=+1$ is motivated by the fact that, in the absence of
rotation, it is a necessary condition for BH scalarization.
However, the last observation about $\mathcal{G}_\note{Kerr}$ suggests that BH
spin might be able to induce scalarization when $\epsilon=-1$. Indeed, it was
shown recently in Ref.~\cite{Dima:2020yac} (see also~\cite{Hod:2020jjy,Doneva:2020nbb}
for follow-up studies) that Kerr BHs are
tachyonically unstable for $f(\phi)=\epsilon\phi^2/2$ and $\epsilon=-1$, once $j$
exceeds a certain threshold (which is above $j=0.5$).
Since this tachyonic instability is the hallmark of spontaneous scalarization,
one expects theories in this class to exhibit a remarkable property:
BHs develop scalar hair only when they spin fast enough.

The approach of Ref.~\cite{Dima:2020yac}, however, does not provide concrete
evidence that these hairy BHs exist. As it focuses on the linearized
equations, it captures only the onset of the tachyonic instability, and it cannot
make conclusive statements about its end point.
In this Letter we instead solve the full field equations numerically to
generate stationary, rotating, asymptotically flat BH solutions.
We show that slowly rotating BHs can only be described by the Kerr solution, as
in GR, whereas, rapidly rotating ones, can indeed have scalar hair.
This is fully consistent with the expectations of Ref.~\cite{Dima:2020yac} and
a clear demonstration that rotation can induce scalar hair if a scalar field
exhibits suitable coupling to curvature.

\bfi{Nonlinear spin-induced scalarized BHs.}
We work with the coupling of Eq.~\eqref{coupling} and $\epsilon=-1$.
At the linear level, this theory coincides with the model studied
in~\cite{Dima:2020yac}, but the end state of the instability, which is our
focus, is sensitive to the nonlinear completion of the theory.
We use the exponential model mostly to facilitate a comparison between our results
and  those of Ref.~\cite{Cunha:2019dwb}, which studied the case $\epsilon=+1$.
We stress that other couplings $f(\phi)$ could have been chosen, including the
quadratic model $f(\phi) = \epsilon  \phi^2 / 2$ or the
effective-field-theory-inspired model of~\cite{Macedo:2019sem}.
We expect all these models to also exhibit the spin-induced spontaneous
BHs scalarization effect, although the nonlinear solutions will have
different properties \cite{Blazquez-Salcedo:2018jnn,Silva:2018qhn,Macedo:2019sem}.

To find these solutions, we work with the ansatz~\cite{Cunha:2019dwb}
\begin{align} \label{eq:ansatz}
\rd s^2 &= - e^{2F_0} N \rd t^2
+ e^{2F_1}\left(N^{-1} \rd r^2
+ r^2 \rd \theta^2\right)
\nonumber \\
&\quad + e^{2F_2}r^2 \sin^2\theta (\rd \varphi-W \rd t)^2 \ ,
\end{align}
where $N \equiv 1 - r_H / r$ and $r=r_H>0$ is the horizon location~\footnote{Only
nonextremal solutions can be studied within the metric ansatz of Eq.~\eqref{eq:ansatz}.}.
The metric functions $F_i$,~$W$ ($i=0,1,2$) and the scalar field $\phi$ depend
on $r,\,\theta$ only.
Asymptotic flatness requires
$\lim_{r\rightarrow \infty}{F_i}=\lim_{r\rightarrow \infty}{W}=\lim_{r\rightarrow \infty}{\phi}=0.$
Axial symmetry and regularity impose the boundary conditions
$ \partial_\theta F_i = \partial_\theta W = \partial_\theta \phi = 0$
on the symmetry axis ($\theta=0,\pi$).
Additionally, the absence of conical singularities implies that $F_1=F_2$ on
the symmetry axis.
The horizon boundary conditions are
$
\partial_x F_i \big|_{r=r_H}= \partial_x \phi  \big|_{r=r_H} =  0$ and $W \big|_{r=r_H}=\Omega_H$,
where, for  convenience, we have introduced a new radial coordinate
$x\equiv(r^2-r_H^2)^{1/2}$. Here $\Omega_H>0$ is the constant horizon angular
velocity. Some details on the numerical scheme used to find the solutions with
these boundary conditions are given in the Supplemental Material,
Sec.~\ref{sm:numethods}~\footnote{See Supplemental Material, at the end of this paper, for further details on the theory's field equations, numerical methods and physical properties
of some of the solutions, which includes Ref.~\cite{schoen}.}

Most of the quantities of interest are encapsulated in the metric functions evaluated either at the
horizon or at infinity.
Consider first horizon quantities. The Hawking temperature is
$T_H={\kappa}/({2\pi})$, where $\kappa$ is the surface gravity defined as
$\kappa^2 \equiv -(1/2)(\nabla_\alpha \xi_\beta)(\nabla^\alpha \xi^\beta)|_{r_H}$
and $\xi \equiv \partial_t + \Omega_H \, \partial_\varphi$ is the
horizon null generator. The area of the spatial sections of the event horizon
is $A_H$.
Explicitly,
\begin{align}
T_H &= (4\pi r_H)^{-1} \cdot e^{F_0(r_H,\theta)-F_1(r_H,\theta)}\,,
\\
A_H &= 2\pi r_H^2 \int_0^\pi \rd \theta \sin \theta~e^{F_1(r_H,\theta)+F_2(r_H,\theta)}\,.
\end{align}

Now consider the asymptotic quantities. The ADM mass $M$ and the angular
momentum $J$ are read off from the asymptotic behavior of the metric functions:
$g_{tt} \simeq -1 + 2M/r$
and
$g_{\varphi t} \simeq - 2J\sin^2\theta / r$.
All solutions reported in this Letter possess also a scalar ``charge'' $Q_s$,
which is found from the scalar field's far-field asymptotic $\phi \simeq - Q_s / r$.
This ``charge"  does not have an associated conservation law, and it
is secondary in the nomenclature of Refs.~\cite{Coleman:1991ku,Kanti:1995vq,Herdeiro:2015waa}.
For all solutions here, both the metric functions and the scalar field are
even parity, i.e., invariant with respect to the transformation
$\theta \to \pi-\theta$. More general solutions, in particular with odd
parity, exist. Typically these are excited states and  unstable, which
justifies our focus  on the  even  parity  sector, corresponding to the
fundamental solutions~\cite{Kunz:2019bhm}.

As in the $\epsilon=+1$~\cite{Cunha:2019dwb} case, the solutions satisfy a
Smarr-type law, and their entropy $S$  has a correction to  the
Bekenstein-Hawking entropy computed  from Wald's formalism~\cite{Wald:1993nt}.
It reads $S=S_E+S_{\rm sGB}$, where
\begin{equation}
\label{S}
S_E=\frac{A_H}{4} \ , \qquad S_{\rm sGB}=  \frac{\lambda^2}{2} \int_{H} \rd^2 x \, \sqrt{h}f(\phi) R^{(2)} \ ,
\end{equation}
with $R^{(2)}$ denoting the Ricci scalar of the metric $h_{ij}$ which is
induced on the spatial sections of the horizon, denoted as $H$.
In the following, we shall use the dimensionless (or reduced) area $a_H \equiv
A_H/(16\pi M^2)$, spin $j=J/M^2$, temperature $t_H\equiv 8\pi T_H M$, and entropy $s \equiv
{S}/({4\pi M^2})$.

\begin{figure}[h!]
\includegraphics[width=\columnwidth]{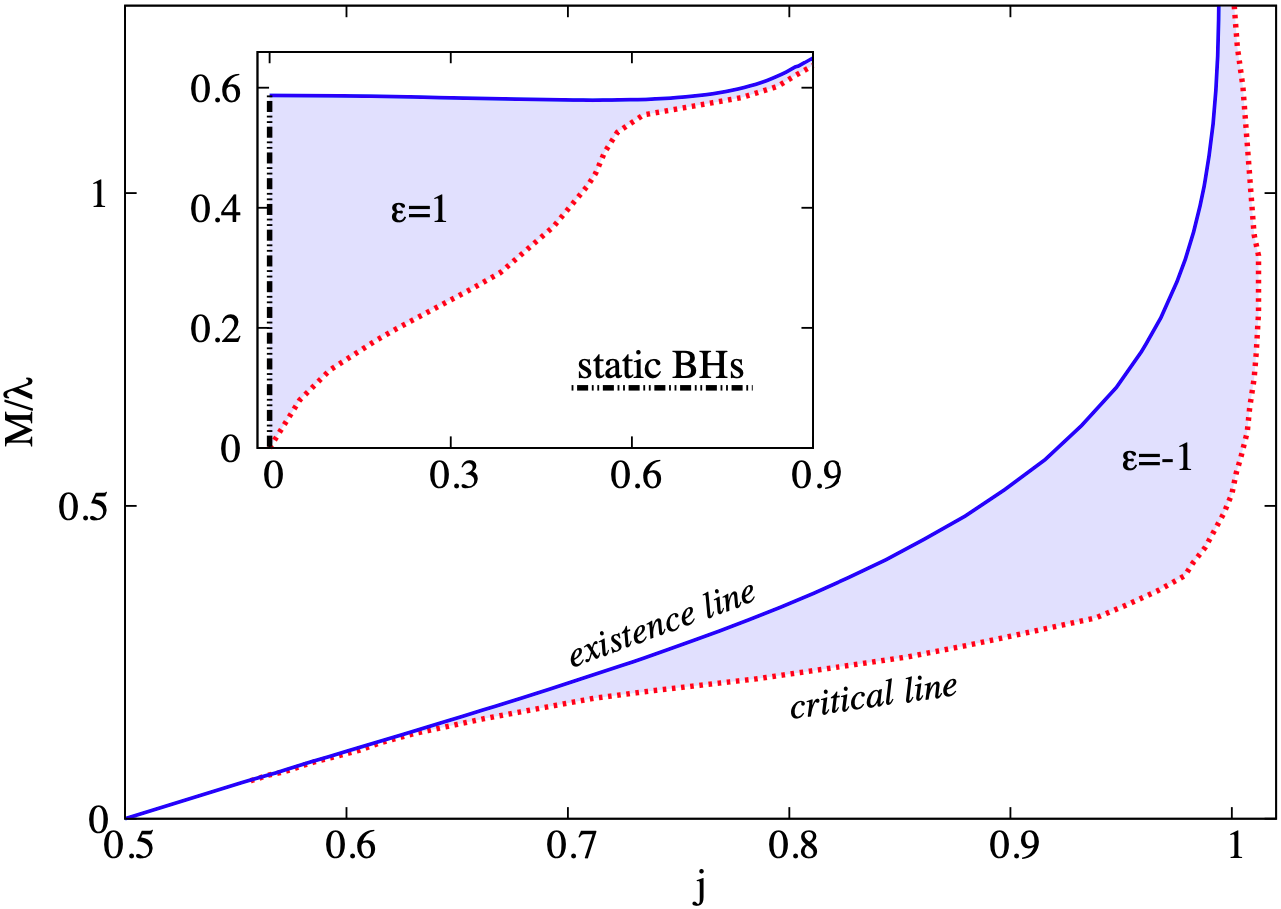}
\includegraphics[width=\columnwidth]{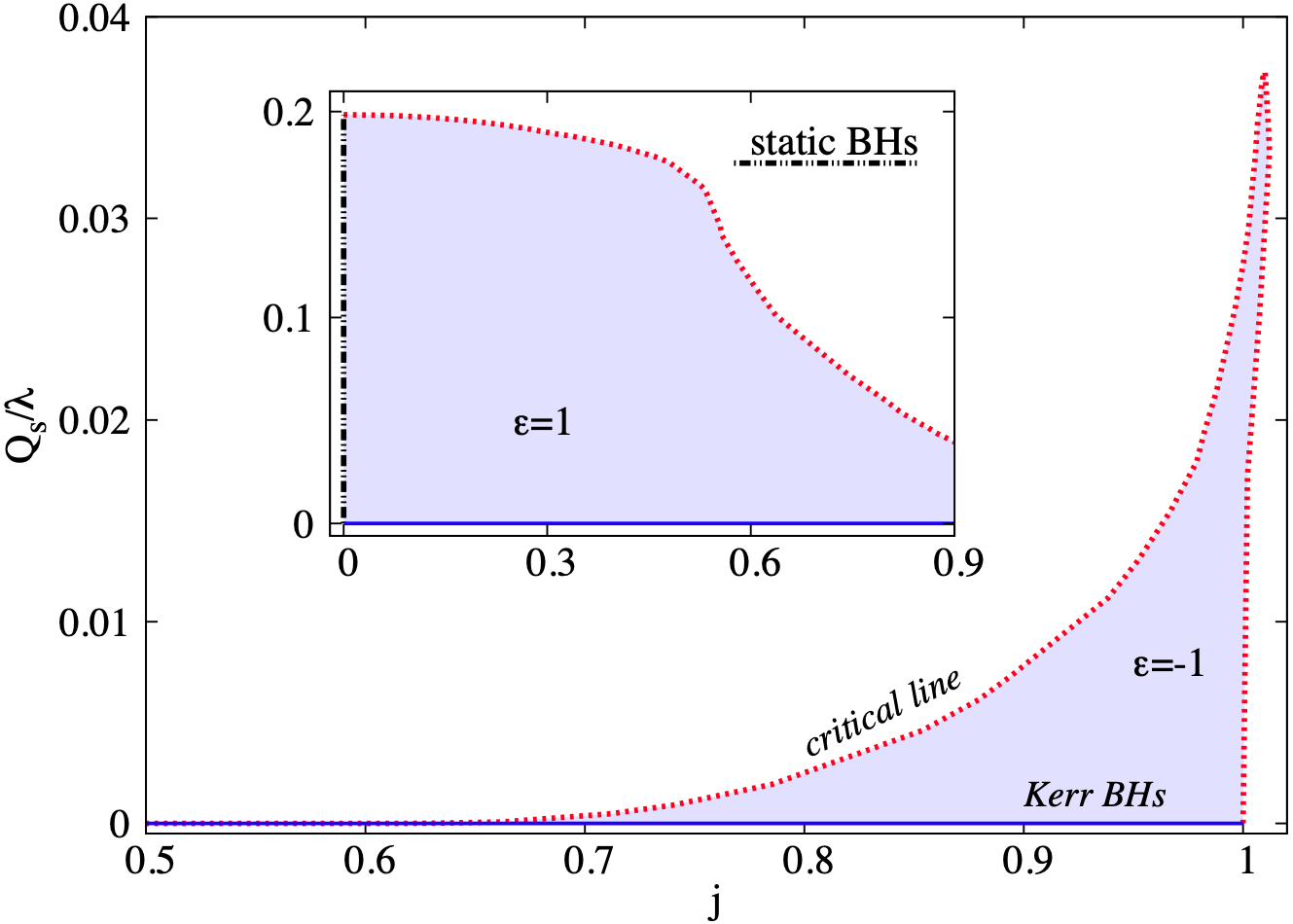}
\caption{ADM mass $M$ (top panel) and scalar charge $Q_s$ (bottom panel),  both in units of
$\lambda$, as
functions of the dimensionless spin $j$ of spinning scalarized BHs. Here and in Fig.~\ref{fig2}, the main panels (insets) correspond to
$\epsilon=-1$ ($\epsilon=+1$).}
\label{fig1}
\end{figure}

\begin{figure}[h!]
\includegraphics[width=\columnwidth]{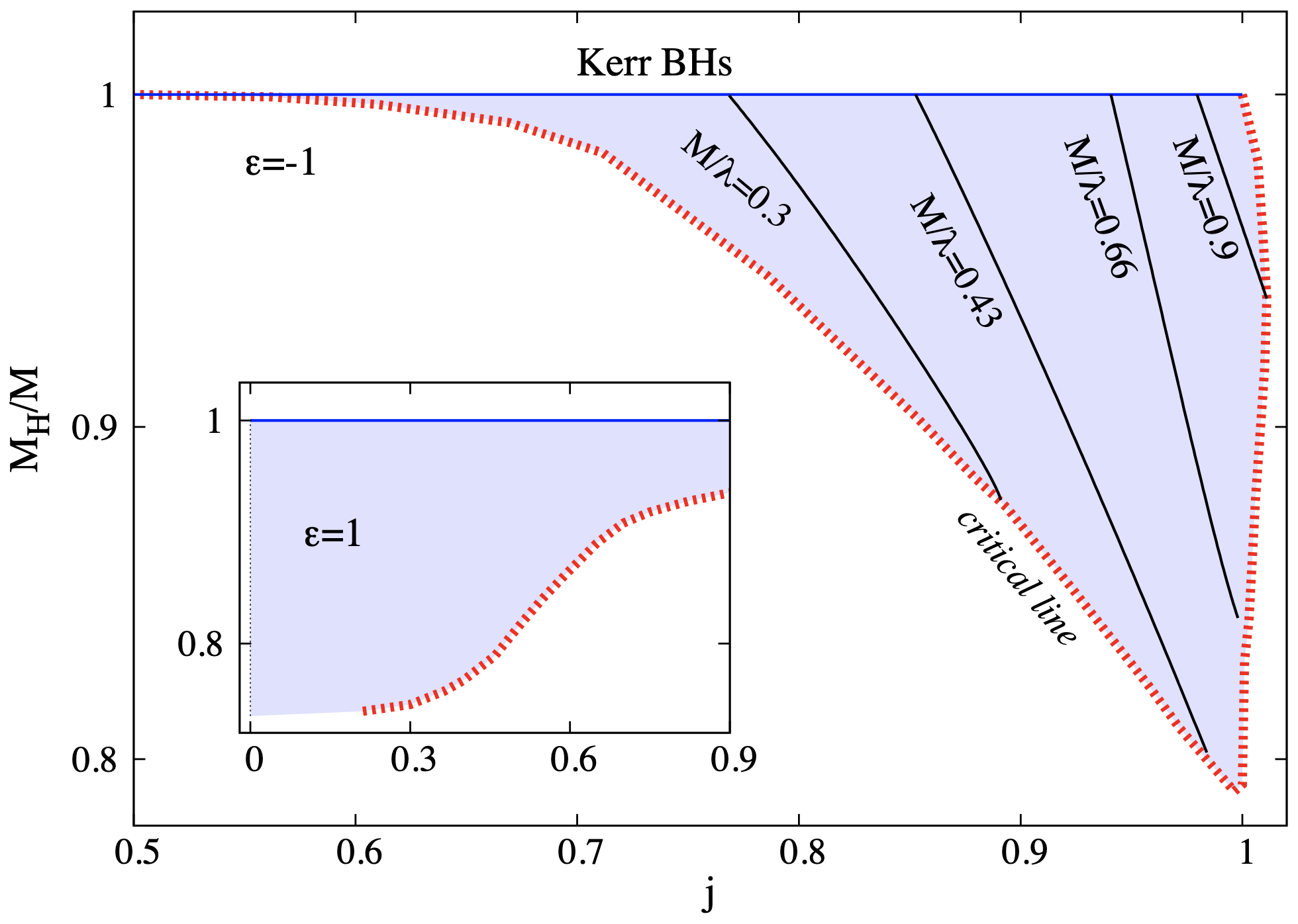}
\includegraphics[width=\columnwidth]{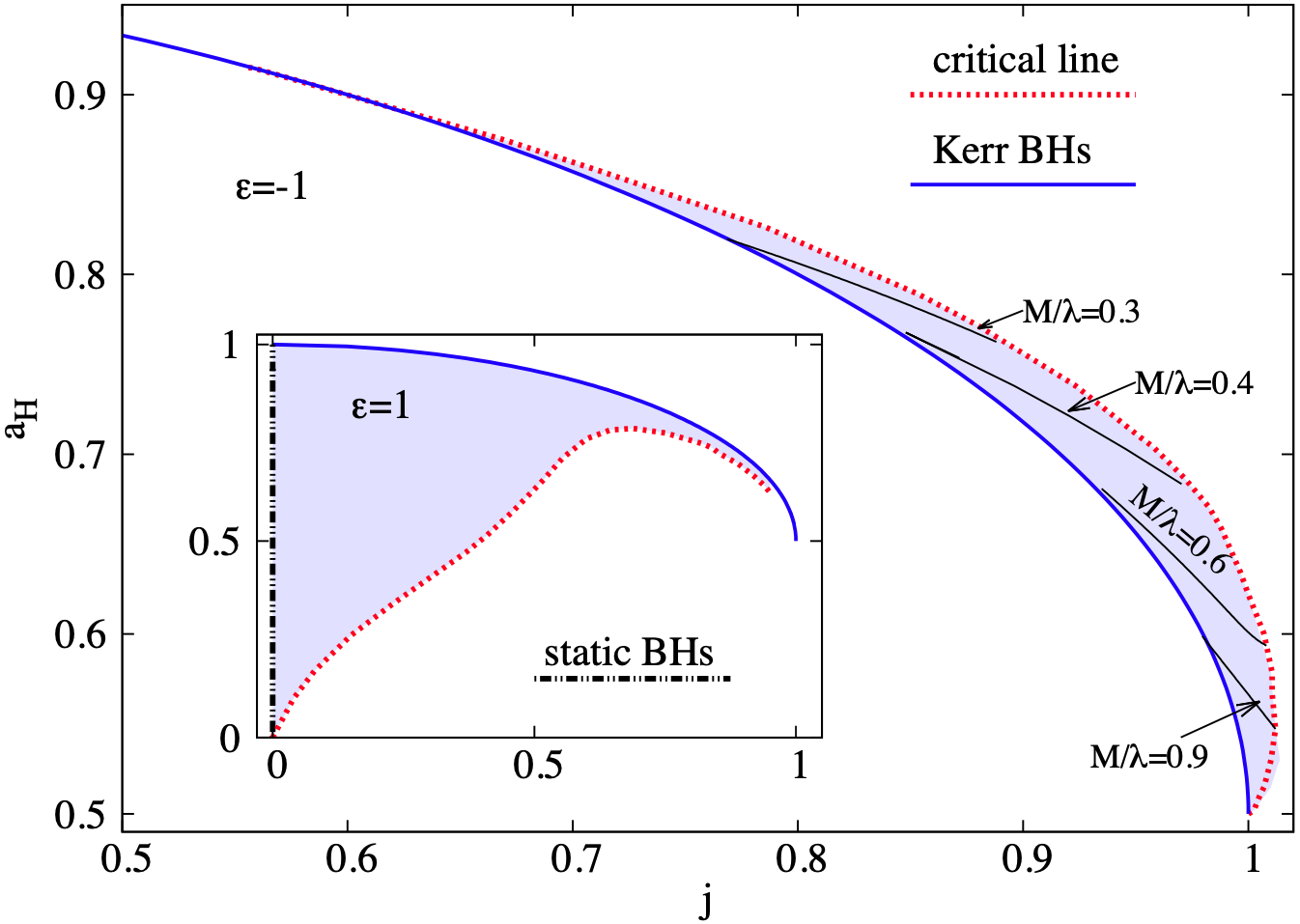}
\includegraphics[width=\columnwidth]{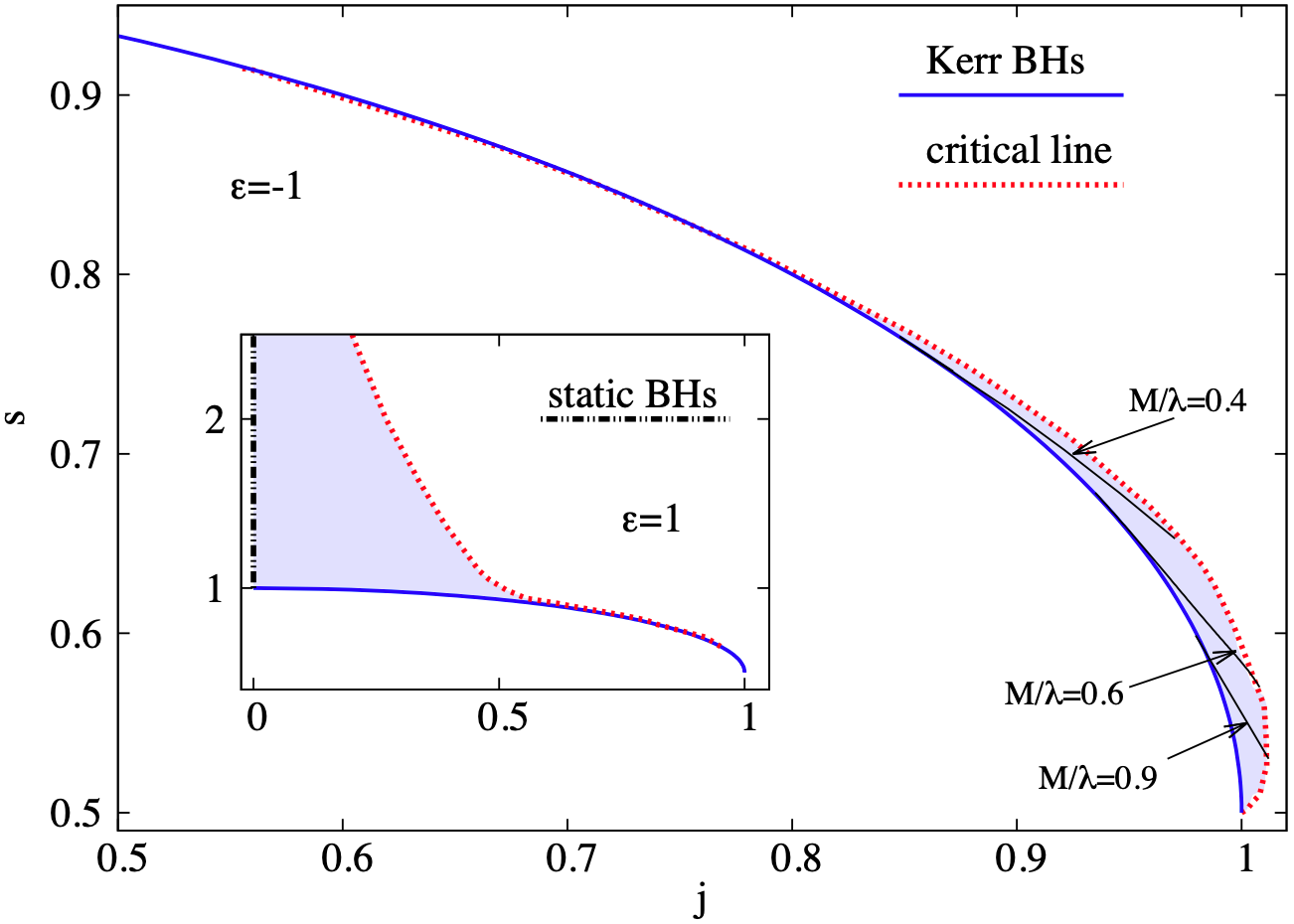}
\caption{Horizon mass over ADM mass $M_H/M$ ratio (top panel), reduced horizon area
$a_H$ (middle panel), and reduced entropy $s$ (bottom panel) as functions of the
dimensionless spin $j$.}
\label{fig2}
\end{figure}

\bfi{Properties of the solutions.}
We have performed a thorough numerical exploration of the parameter space to
examine the domain of existence and the physical properties of the spinning
scalarized BHs. This domain of existence is represented in all panels of
Figs.~\ref{fig1} and \ref{fig2} by the darker shaded area, being obtained by
extrapolating to the continuum the results from a set of around 1000 numerical solutions.

Figure~\ref{fig1} (top panel) exhibits an overview  of the domain of existence
in an $M/\lambda$~vs.~$j$ plot. Consider first the limits of the domain of
existence, which in fact appear in all panels of the subsequent figures.
For $\epsilon=-1$, the domain is bounded by two sets of solutions: (i) the
``existence line,'' which corresponds to the bifurcation edge from the Kerr
family (see the solid blue line in Figs.~\ref{fig1}~and~\ref{fig2}), and (ii) the set
of ``critical solutions'' (dotted red lines in
Figs.~\ref{fig1}~and~\ref{fig2})~\footnote{The critical solutions interpolating
between the scalarized BH with maximal $j$ and extremal Kerr BHs have a small
Hawking temperature, which suggests the possible existence of a (small) set of
extremal BHs. However, this issue has proved to be numerically difficult to
explore.}.
A third boundary exists when $\epsilon=+1$, the ``static configurations''
\cite{Cunha:2019dwb,Collodel:2019kkx} (dashed-dotted black lines in the insets of Figs.~\ref{fig1}~and~\ref{fig2}).

The existence line is universal for any coupling function allowing for
scalarization. In principle, this particular set of solutions can be found by
solving the scalar field equation (as  a test field) on the Kerr background.
In our approach, however, the existence line is found as the limiting
configuration wherein $\phi\to 0$, when varying  $r_H,\Omega_H$ for fixed
$\lambda$. Some quantitative details on the existence line are given in
the Supplemental Material, Sec.~\ref{sm:existence_line}.

The set of critical solutions is model dependent. The numerical process
fails to converge as this set of configurations is approached. Typically,
neither a singular behavior nor a deterioration of the numerical accuracy in
the vicinity of this set was observed.
The existence of such critical solutions in fairly commonplace in sGB models, both for spherical \cite{Kanti:1995vq,Sotiriou:2013qea,Sotiriou:2014pfa} and rotating \cite{Kleihaus:2011tg,Kleihaus:2015aje} hairy BHs.
An explanation can be traced back to the fact that the radicand of a square
root in the horizon expansion of the scalar field vanishes as the critical set
is approached (see e.g., Appendix A in~\cite{Kleihaus:2015aje} or Sec.~5.1 in~\cite{Delgado:2020rev}).
As such, a consistent near horizon expansion of the solution ceases to exist,
indicating that a solution that is regular there does not exist.

From Fig.~\ref{fig1} (top panel), we see that $\epsilon= -1$ solutions exist
for a range of values of the dimensionless spin $0.5<j\lesssim 1$.
Concerning the lower limit, the minimum $j$ value retrieved along the existence
line with our procedure is $j\simeq 0.55$.
This is compatible with the fact that the spin-induced scalarization
instability of Kerr can only exist for $j>0.5$ and also with the results
in~\cite{Dima:2020yac,Hod:2020jjy,Doneva:2020nbb}.
Concerning the upper limit, within the dataset collected, the maximal value of
$j$ for the scalarized BHs slightly exceeds unity: $ j_{\rm max} \sim 1.01$.
This means that scalarized BHs in this model exhibit small violations of the
Kerr bound.
This $j$ range contrasts with the $\epsilon=+1$ case (inset), which extends down
to $j=0$. As a similar trend for both $\epsilon=\pm 1$, for a given $\lambda$, the
maximal allowed scalarized BH mass increases with $j$ (for $\epsilon=+1$ this
holds for sufficiently high angular momentum).
This assertion relies on the shape of the existence line and it is thus
universal for any coupling function $f(\phi)$ allowing for scalarization.

Now let us examine some of the physical properties of the solution.  First, how
much scalar ``hair'' do the scalarized solutions posses? Several quantities can
be used to address this question.
In Fig.~\ref{fig1} (bottom panel), the  scalar charge (in units of $\lambda$) is
represented against $j$.
In the $\epsilon=+1$ case, this charge is maximized for static $j=0$ solutions.
For the $\epsilon=-1$ case, it is maximized (within numerical accuracy) for
$j\sim j_{\rm max}$, corresponding to $Q_s/\lambda\sim  0.038$.
Comparing it to Fig.~\ref{fig1} (top panel), we conclude that the maximal $Q_s$
occurs for $M/\lambda\simeq 0.9$.

A comparison between Fig.~\ref{fig1} (bottom panel) and Fig.~\ref{fig2} (top panel) also
reveals  that $Q_s$ is no faithful measure  of the fraction of the mass stored
in the BH (and hence the fraction stored in the scalar field), as these two
quantities are not extremized for the same $M/\lambda$~\footnote{The horizon
mass $M_{\rm H}$ is computed as a Komar integral.}.
In this respect, Fig.~\ref{fig2} (top panel) shows that a significant part of
the total mass is stored outside the horizon. For $\epsilon=+1$, this fraction
obeys $M_{H}/M\gtrsim 0.735$, whereas for $\epsilon=-1$, $M_{H}/M\gtrsim 0.79$.
This suggests significant differences in some phenomenological properties,
e.g., geodesic motion and BH shadows, may exist with respect to comparable
Kerr BHs. These difference, moreover, should be enhanced for larger $j$ up to
near the maximal $j$.

An important distinction between the $\epsilon=\pm 1$ models concerns the
horizon area. Figure~\ref{fig2} (middle panel) shows that, for the same $j$,
$a_H$ is maximized (minimized) by the Kerr solution for $\epsilon=+1$
($\epsilon=-1$).
In this sense, spin-induced scalarized BHs are larger than Kerr, whereas they
are smaller in the gravitoelectric ($j \leqslant 0.5$) led scalarization.
Yet, in  both cases, they are entropically favored over Kerr [see Fig.~\ref{fig2} (bottom panel)].
This is partly explained by the fact that the correction to the GR BH entropy
depends on the sign of $f(\phi)$, cf. Eq.~\eqref{S}.
We remark, however, that the entropic preference for the same $M, J$ in axial
symmetry may be less significant for the dynamical preference than in spherical
symmetry, as gravitational radiation can be emitted during the process of
scalarization for the former but not the latter.

\bfi{Conclusions.}
We have solved the full field equations to generate solutions that describe
stationary, rotating BHs in an illustrative model [cf. Eq.~\eqref{coupling}]
that exhibits the spin-induced tachyonic instability found in
Ref.~\cite{Dima:2020yac}.
Our results clearly demonstrate that slowly spinning stationary BHs in this
model are described by the Kerr solution, whereas rapidly spinning ones exhibit
scalar hair.
The transition between the two classes of solutions takes place right on the
threshold of the tachyonic instability found in Ref.~\cite{Dima:2020yac}.
Hence, the hairy solutions are expected to be end states of spin-induced BH
scalarization.

Spin-induced scalarization raises the exciting possibility that astrophysical
BHs will defy the Kerr hypothesis only for large spins, which merits further
investigation. We have already established that the scalarized BH solutions
are entropically preferred in the regime of the tachyonic instability, but it
would be interesting to study their stability properties.
It would also be important to follow dynamically the development of the
tachyonic instability found in Ref.~\cite{Dima:2020yac}, track the formation of
scalar hair, and verify explicitly that the solutions found here are the end
points of this instability.
This has been achieved in simpler BH scalarization
scenarios~\cite{Herdeiro:2018wub}, but it is particularly challenging when one
has a coupling with the Gauss-Bonnet invariant, although significant progress
has recently been made in modeling nonlinear time-domain evolutions in these
theories~\cite{Benkel:2016kcq,Benkel:2016rlz,Witek:2018dmd,Okounkova:2019zep,Okounkova:2020rqw,Ripley:2019aqj,Ripley:2019irj,Ripley:2020vpk,Witek:2020uzz,Julie:2020vov}.
Finally, the astrophysical phenomenology and implications of the scalarized BHs
reported herein is missing and our results hold the promise of non-negligible
deviations from the Kerr phenomenology.

\bfi{Acknowledgments.}
We thank H. Witek for discussions.
C.H. and E.R. acknowledge support by the Center for Research and Development in
Mathematics and Applications (CIDMA) through the Portuguese Foundation for
Science and Technology (FCT - Funda\c{c}\~{a}o para a Ci\^encia e a Tecnologia),
references  UIDB/04106/2020  and  UIDP/04106/2020, and by national funds
(OE), through FCT, I.P., in the scope of the framework contract foreseen in
the numbers 4, 5, and 6 of the article 23 of the Decree-Law 57/2016 of August
29, changed by Law 57/2017 of July 19.
We acknowledge support from the projects PTDC/FIS-OUT/28407/2017 and
CERN/FIS-PAR/0027/2019. This work has further been supported by the
European Union Horizon 2020 research and innovation (RISE) program
H2020-MSCA-RISE-2017 Grant No.~FunFiCO-777740.
H.O.S. and N.Y. are supported by NASA Grants Nos.~NNX16AB98G,
~80NSSC17M0041, and~80NSSC18K1352 and NSF Award No.~1759615.
T.P.S. acknowledges partial support from the STFC Consolidated Grant
No.~ST/P000703/1.
We also acknowledge networking support by the COST Action
GWverse Grant No.~CA16104.

\vspace{0.2cm}
\bfi{Note added.}~Recently, we became aware of a manuscript which independently
derives similar results for a different coupling function~\cite{Berti:2020kgk}.

\bibliography{master}

\clearpage
\newpage
\begin{center}
\textbf{-- Supplemental Material --}
\end{center}

\section{Field equations.}
\label{sm:field_eqs}
%
The field equations obtained from the action Eq.~\eqref{eq:gbaction} are:
\begin{align}
R_{\mu\nu} - \frac{1}{2} g_{\mu\nu} R = \frac{1}{2} T_{\mu\nu} \,,
\quad
\Box \phi = - \frac{\lambda^2}{4} f'(\phi) \mathcal{G} \,,
\end{align}
where $T_{\mu\nu}$ is an effective energy-momentum tensor
\begin{equation}
T_{\mu\nu} = 4 T^{(\phi)}_{\mu\nu} - \lambda^2 T^{(\mathcal{G})}_{\mu\nu}  \,,
\end{equation}
composed of the scalar field energy-momentum tensor
\begin{equation}
T^{(\phi)}_{\mu\nu} = \nabla_{\mu}\phi \nabla_{\nu}\phi
- \frac{1}{2} g_{\mu\nu} \nabla_{\alpha}\phi \nabla^{\alpha}\phi \,,
\end{equation}
and another due to the coupling between scalar field $\phi$ and Gauss-Bonnet invariant $\mathcal{G}$
due to the coupling $\lambda^2 f(\phi)$
\begin{equation}
T^{(\mathcal{G})}_{\mu\nu} =
2 g_{\alpha(\mu} g_{\nu)\beta} \epsilon^{\gamma \beta \delta \kappa} \nabla_{\lambda}\left[^{\ast}R^{\alpha\lambda}{}_{\delta\kappa} f'(\phi) \nabla_{\gamma} \phi \right]
\,,
\end{equation}
where $^{\ast}R^{\mu\nu}{}_{\rho\sigma} = \epsilon^{\mu\nu\alpha\beta}\, R_{\alpha\beta\rho\sigma}$ is the dual Riemann tensor,
$\epsilon^{\mu\nu\rho\sigma}$ is the totally anti-symmetric Levi-Civita pseudo-tensor and, as defined in the main text, $f'(\phi) = {\rm d} f(\phi) / {\rm d} \phi$.

\section{Numerical methods.}
\label{sm:numethods}
The field equations reduce to a set of five coupled non-linear elliptic partial
differential equations for the functions ${\cal F}_a =(F_0, F_1, F_2, W;
\phi)$, which are found by plugging the ansatz \eqref{eq:ansatz} together with
$\phi=\phi(r,\theta)$ into the field equations derived from Eq.~\eqref{eq:gbaction} with  the coupling of Eq.~\eqref{coupling}.
These equations have been solved  subject to the boundary conditions introduced in  the main text.

The numerical treatment can be summarized as follows. The domain of
integration is restricted to the region outside the horizon, $r\geqslant r_H$.
A new (compactified) radial variable $\bar x=x/(1+x)$ is introduced, which maps
the semi--infinite region $[0,\infty)$ to the finite region $[0,1]$, where
$x\equiv \sqrt{r^2-r_H^2}$ and $r$ is the radial variable in the line element
\eqref{eq:ansatz}.
Next, the equations for ${\cal F}_a$ are discretized on a grid in $\bar x$ and
$\theta$, which covers the integration region
$0\leqslant \bar x \leqslant 1$ and $0\leqslant \theta \leqslant \pi/2$.
Most of the results in these notes have been found for an equidistant grid with
$250 \times 30$ points.

All numerical calculations are performed by using a professional package
\cite{schoen}, which employs a  Newton-Raphson method.  This code uses  the
finite difference method, providing also an error estimate for each unknown
function.  The numerical error for the solutions reported in this work is
estimated to be typically $<10^{-3}$.  In deriving the equations for
${\cal F}_a$ and in the analysis of the numerical output we have used mainly
\textsc{mathematica}.

After fixing the model, in particular $f(\phi),\epsilon$, the solutions space
is scanned by using the following input parameters: the event horizon radius
$r_H$, the  horizon angular velocity $\Omega_H$ and the coupling constant
$\lambda$ (which specifies the scale in the action).
We fix $\lambda$ and construct the domain of existence by varying both $r_H$
and $\Omega_H$.

\section{First law of BH thermodynamics.}
\label{sm:firstlaw}
%
In scalar-Gauss-Bonnet gravity, BHs satisfy the relation
\begin{equation}
\rd M=T_H \, \rd S +\Omega_H \,\rd J\,,
\end{equation}
where the contribution from the scalar field is not explicit. We used this
formula to validate our numerical integration.

\section{The existence line.}
\label{sm:existence_line}
\begin{figure*}[htb]
\includegraphics[width=\columnwidth]{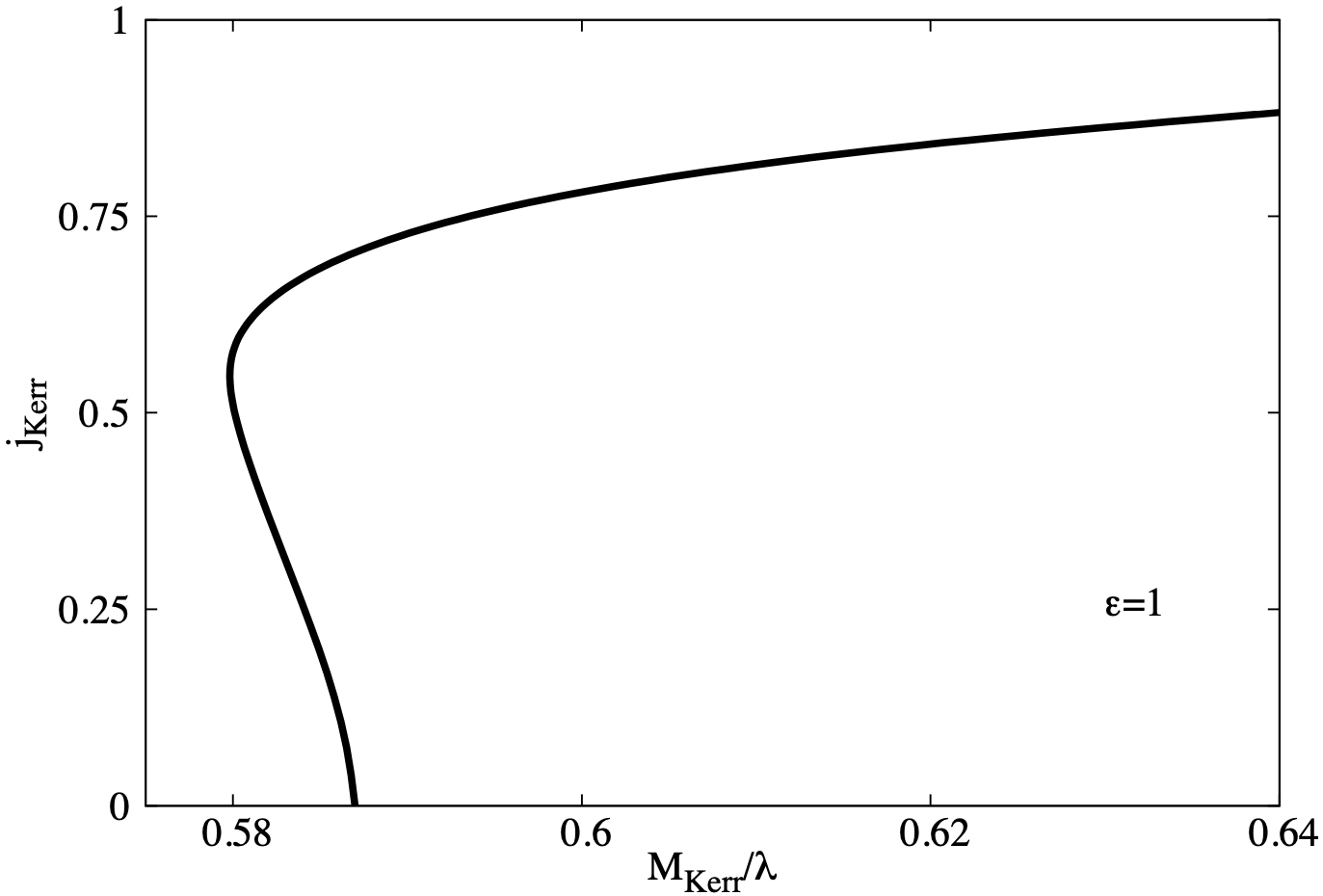}
\includegraphics[width=0.97\columnwidth]{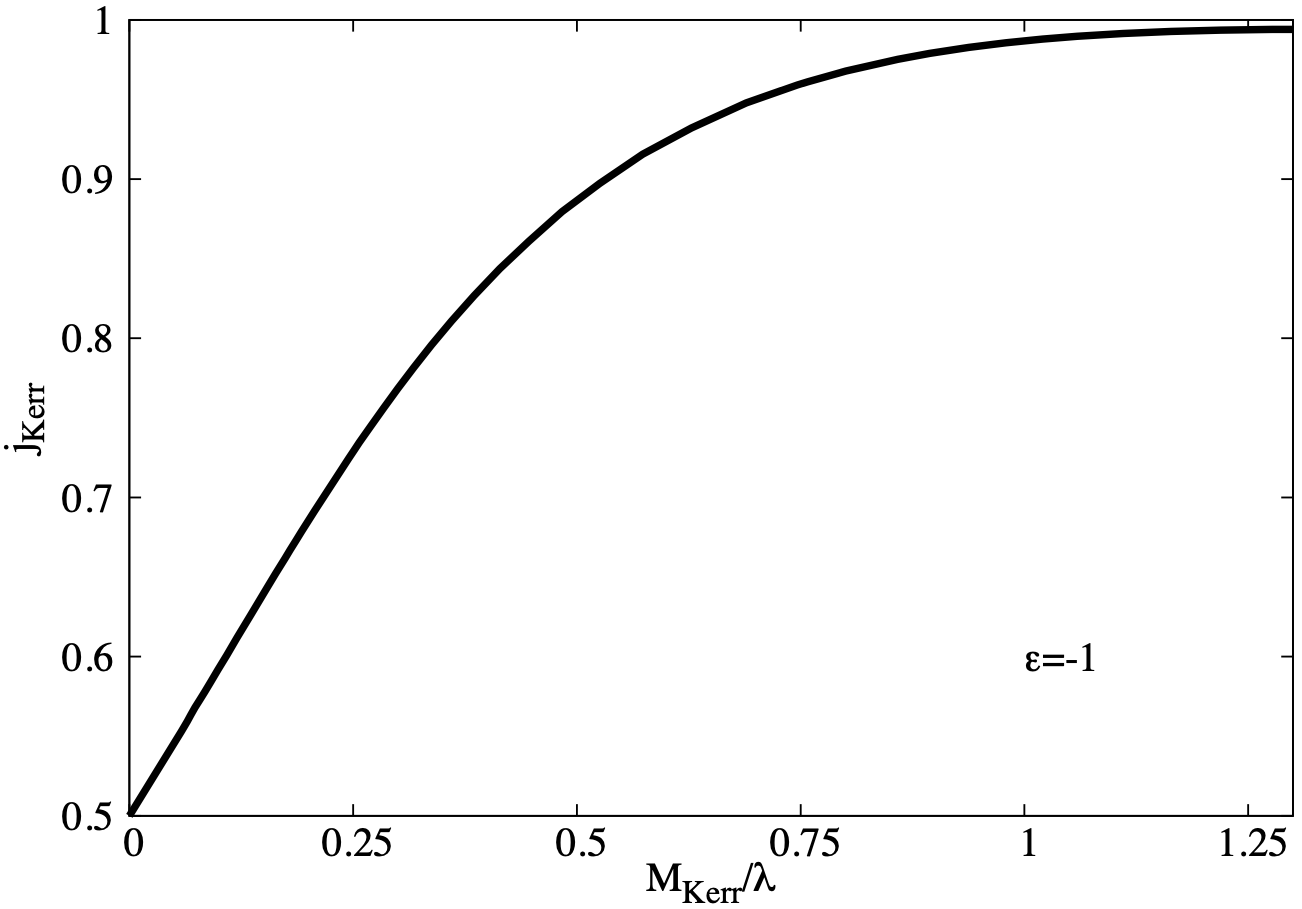}
\caption{Reduced angular momentum $j$,
of the set of Kerr BHs along the existence line,
$vs.$  $M$, in units of $\lambda$.
The lines result from an interpolation of numerical points.
The left and right panels correspond to
$\epsilon=+1$
and
$\epsilon=-1$, respectively.
}
\label{existence-line}
\end{figure*}

%
The existence line is exhibited in Fig.~\ref{existence-line}, where $j$ is shown as a function
of the Kerr BH mass (in units of $\lambda)$ for both $\epsilon=\pm 1$.
In the case $\epsilon=-1$, one notices that the instability  occurs for $1/2<j<1$;
to be precise, the last data points correspond to $j\simeq 0.55$ and $j\simeq 0.994$, respectively.
Some data points along the existence line are given in Table I for both $\epsilon=\pm 1$.

 \begin{table}[htb!]
 \begin{tabular}{c c c | c c c }
  \hline
  \hline
  $M/\lambda$ &  $J/\lambda^2$ &  $j$  & $M/\lambda$ &  $J/\lambda^2$ &  $j$    \\
  \hline
  $0.587$        &  $0.000$        &  $ 0.000 $  & $0.061$        &  $0.002$        &  $ 0.556 $   \\
  $0.586$        &  $0.035$        &  $ 0.103 $  & $0.109$        &  $0.007$        &  $ 0.601 $   \\
  $0.584$        &  $0.085$        &  $ 0.249 $  & $0.164$        &  $0.017$        &  $ 0.652 $   \\
  $0.581$        &  $0.138$        &  $ 0.408 $  & $0.192$        &  $0.025$        &  $ 0.679 $   \\
  $0.579$        &  $0.185$        &  $ 0.552 $  & $0.257$        &  $0.048$        &  $ 0.734 $   \\
  $0.580$        &  $0.205$        &  $ 0.610 $  & $0.318$        &  $0.079$        &  $ 0.781 $   \\
  $0.586$        &  $0.240$        &  $ 0.700 $  & $0.414$        &  $0.144$        &  $ 0.844 $   \\
  $0.594$        &  $0.266$        &  $ 0.756 $  & $0.628$        &  $0.368$        &  $ 0.932 $   \\
  $0.606$        &  $0.295$        &  $ 0.805 $  & $0.801$        &  $0.620$        &  $ 0.968 $   \\
  $0.619$        &  $0.321$        &  $ 0.840 $  & $0.979$        &  $0.945$        &  $ 0.986 $   \\
  $0.635$        &  $0.351$        &  $ 0.870 $  & $1.164$        &  $1.344$        &  $ 0.993 $   \\
\hline
\hline
\end{tabular}
\vspace{0.3cm}
\caption{The ADM mass $M$, angular momentum $J$  (in units of $\lambda$) and dimensionless spin parameter $j$ of some data points on the existence line, for $\epsilon=+1$ (left) $\epsilon=-1$ (right).}
\end{table}

\end{document}